\documentclass[12pt]{article}
\usepackage{jhep-mod}
\usepackage{bm}
\usepackage{amssymb,amsmath,amsthm}
\usepackage{mathrsfs}
\usepackage[utf8]{inputenc}
\usepackage{enumerate}
\usepackage{appendix}
\usepackage{graphicx}
\usepackage{float}
\usepackage{tikz}
\usepackage{setspace}
\usepackage{cancel}
\definecolor{purple}{rgb}{1,0,1}
\definecolor{lime}{HTML}{A6CE39} 

\newcommand{\blue}[1]{{\slshape\color{blue} #1}}

\definecolor{lime}{HTML}{A6CE39}
\newcommand{\orcidicon}{%
	\begin{tikzpicture}
	\draw[lime, fill=lime] (0,0) 
		circle [radius=0.16] 
		node[white] {{\fontfamily{qag}\selectfont \tiny ID}};
	\draw[white, fill=white] (-0.0625,0.095) 
		circle [radius=0.007];
	\end{tikzpicture}
	\hspace{-5mm}
}
\newcommand\orcidAlex{{\href{https://orcid.org/0000-0002-1763-3563}{\orcidicon}}}
\newcommand\orcidPrado{{\href{https://orcid.org/0000-0001-8073-4896}{\orcidicon}}}
\newcommand\orcidMatt{{\href{https://orcid.org/0000-0003-1088-6485}{\orcidicon}}}
\begin{document}

\title{\vspace{-25pt}\huge{Vaidya spacetimes, black-bounces, \\ and traversable wormholes}}

\author{Alex Simpson$^\dagger$\orcidAlex, Prado Mart\'in--Moruno$^\ddagger$\orcidPrado, and Matt Visser$^1$\orcidMatt}

\affiliation{$^\dagger$School of Mathematics and Statistics, Victoria University of Wellington, \\
\null\qquad PO Box 600, Wellington 6140, New Zealand}
\emailAdd{alex.simpson@sms.vuw.ac.nz, matt.visser@sms.vuw.ac.nz}
\affiliation{$^\ddagger$Departamento de F\'isica Te\'orica and IPARCOS, \\ \null\qquad Universidad Complutense de Madrid,
E-28040 Madrid, Spain}
\emailAdd{\null\null\qquad pradomm@ucm.es}
\vspace{10pt}

\abstract{
\null\quad \\
We consider a non-static evolving version of the regular ``black-bounce''/\-traversable wormhole geometry recently introduced in JCAP02(2019)042.
We first re-write the static metric using Eddington--Finkelstein coordinates, 
and then allow the mass parameter $m$ to depend on the null time coordinate (\emph{\`a la Vaidya}). The spacetime metric is
\[
ds^{2}=-\left(1-\frac{2m(w)}{\sqrt{r^{2}+a^{2}}}\right)dw^{2}-(\pm 2 \,dw \,dr)
+\left(r^{2}+a^{2}\right)\left(d\theta^{2}+\sin^{2}\theta \;d\phi^{2}\right).
\]
Here $w=\{u,v\}$ denotes suitably defined $\{outgoing,ingoing\}$ null time coordinates; representing $\{retarded,advanced\}$ time, while, (at least for $a\neq 0$), we allow $r\in(-\infty,+\infty)$.
This spacetime is still simple enough to be tractable, and neatly interpolates between Vaidya spacetime, a black-bounce, and  a traversable wormhole. 
We show how this metric can be used to describe several physical situations of particular interest, including a growing black-bounce, a wormhole to black-bounce transition, and the opposite black-bounce to wormhole transition.

\smallskip
\noindent
{\sc Date:} 12 February 2019; 25 June 2019; \LaTeX-ed \today

\smallskip
\noindent{\sc Keywords}:\\
Vaidya spacetime; regular black hole; black-bounce; null-bounce; traversable wormhole.

\smallskip
\noindent
{\sc Pacs}: 04.20.-q; 04.20.Gz; 04.70.-s; 04.70.Bw

\smallskip
\noindent
{\sc arXiv}:  1902.04232 [gr-qc]

\smallskip
\noindent
{\sc Published}:  Classical and Quantum Gravity {\bf 36  \# 14}   (2019)  145007.

\smallskip
\noindent
{\sc DOI:} 
\url{https://dx.doi.org/10.1088/1361-6382/ab28a5}
}

\maketitle
\def\tr{{\mathrm{tr}}}
\def\diag{{\mathrm{diag}}}
\def\cof{{\mathrm{cof}}}
\def\pdet{{\mathrm{pdet}}}
\parindent0pt
\parskip7pt
\section{Introduction}
\enlargethispage{20pt}
Ever since Bardeen initially proposed the concept of a regular black hole over 50 years ago in 1968~\cite{Bardeen:1968}, this notion has continually attracted significant attention.  See for instance the discussion in references~\cite{Bergmann-Roman,Hayward:2005, Bardeen:2014, Frolov:2014, Frolov:2014b, Frolov:2016, Frolov:2017, Frolov:2018,Cano:2018,Bardeen:2018, regular, beyond, black-bounce}. 
Specifically, in reference~\cite{black-bounce} two of the current authors considered the static spacetime covered by coordinate patches of the form:
\begin{equation}
    ds^{2}=-\left(1-\frac{2m}{\sqrt{r^{2}+a^{2}}}\right)dt^{2}+\frac{dr^{2}}{1-\frac{2m}{\sqrt{r^{2}+a^{2}}}}+\left(r^{2}+a^{2}\right)\left(d\theta^{2}+\sin^{2}\theta d\phi^{2}\right).
\end{equation}
Adjusting the parameter $a$, assuming without loss of generality that $a>0$, and following the analysis of reference~\cite{black-bounce}, this metric represents either: 
\vspace{-3pt}
\begin{enumerate}
\itemsep-1pt
\item 
The ordinary Schwarzschild spacetime ($a=0$); 
\item
A ``black-bounce'' with a one-way spacelike throat  ($a<2m$);
\item
A one-way wormhole with a null throat ($a=2m$), \\
compare with reference~\cite{Cano:2018}; or
\item
A traversable wormhole in the Morris--Thorne sense ($a>2m$), 
see~\cite{Morris:1988a, Morris:1988b, Visser:1989a, Visser:1989b, Lorentzian, Visser:2003, Hochberg:1997, Poisson:1995, Barcelo:2000, Hochberg:1998, Cramer:1994, Visser:1997, Barcelo:1999, Garcia:2011, Boonserm:2018, Lobo:2004}.
\end{enumerate} 
\vspace{-3pt}
In the current article we explore a (relatively) \emph{tractable} way of adding time dependence to this spacetime.

We start by re-writing the static spacetime in Eddington--Finkelstein coordinates using coordinate patches of the form
\begin{equation}
ds^{2}=-\left(1-\frac{2m}{\sqrt{r^{2}+a^{2}}}\right)dw^{2}-( \pm 2 \,dw \,dr)
+\left(r^{2}+a^{2}\right)\left(d\theta^{2}+\sin^{2}\theta \;d\phi^{2}\right).
\end{equation}
Here $w=\{u,v\}$ is a suitably defined $\{outgoing,ingoing\}$ null time coordinate. That is, in the outer asymptotic region $r>\sqrt{\max\{0,4m^2-a^2\}}$ the coordinate $w$ manifestly represents $\{retarded,advanced\}$ time, while in the remaining portion of the chart,  in the $r\leq\sqrt{\max\{0,4m^2-a^2\}}$ region, we continue to use the same nomenclature for the  $w=\{u,v\}$ coordinates. 
Here the upper + sign corresponds to $u$, and the lower $-$ sign corresponds to $v$.
Note that as long as $a\neq0$ we can permit the $r$-coordinate to take negative values, $r\in(-\infty,+\infty)$. Then, when the geometry represents a traversable wormhole, we may naturally extend the region of analysis into the ``other'' universe connected by the wormhole throat at $r=0$.
We might need, and sometimes will need, several coordinate patches of this form to cover the maximally extended spacetime --- see discussion below.

We now invoke a Vaidya like trick~\cite{Vaidya:1951a,Vaidya:1951b,Vaidya:1999a,Vaidya:1999b,Vaidya:1970, Wang:1998, Parikh:1998}, by allowing the mass parameter $m(w)$ to depend on the null time coordinate. That is we consider the spacetime described by the metric
\begin{equation}\label{E:dmetric}
ds^{2}=-\left(1-\frac{2m(w)}{\sqrt{r^{2}+a^{2}}}\right)dw^{2}-(\pm 2 \,dw \,dr)
+\left(r^{2}+a^{2}\right)\left(d\theta^{2}+\sin^{2}\theta \;d\phi^{2}\right).
\end{equation}
When $a\to0$ this is just the standard Vaidya spacetime~\cite{Vaidya:1951a,Vaidya:1951b,Vaidya:1999a,Vaidya:1999b,Vaidya:1970, Wang:1998, Parikh:1998},  (either a ``shining star" or a star accreting a flux of infalling null dust).
This metric can be used either to study the collapse of null dust, or the semiclassical evaporation of black holes.

When the parameter $m(w)\to m$ is a constant we just have the static black-bounce/ traversable wormhole of reference~\cite{black-bounce}. 
The point of \blue{now} introducing time dependence in this precise manner is to keep calculations algebraically tractable; and so provide a simple model of an evolving (either through net evaporation or accretion) regular black hole. Another considerably less tractable option, which will not be explored in this paper, would consist of promoting the parameter $a$ to $a(w)$, with $m$ either kept constant or not.

So it is natural to argue that, on one hand, for an increasing function $m(v)$ crossing the $a/2$ limit, the spacetime metric (\ref{E:dmetric}) describes the conversion of a wormhole into a regular black hole by the accretion of null dust. On the other hand, for a decreasing  function $m(u)$ crossing the $a/2$ limit, the situation will correspond to the evaporation of a regular black hole leaving a wormhole remnant. 
Moreover, this may be related to the more-or-less equivalent process of phantom energy accretion onto black holes, which should, however, be studied considering negative energy and using the ingoing null coordinate $v$. (For related discussion see references~\cite{Babichev:2004yx,Babichev:2004qp,MartinMoruno:2006mi,GonzalezDiaz:2007gt,Martin-Moruno:2007,Madrid:2010} and \cite{Lobo:2014,Lobo:2013}.) Finally, it is worth noticing that one can describe the transmutation of a regular black hole into a wormhole and \emph{vice versa} in this classical description only because the black hole is regular and, therefore, there is no topology change. 
It should be noted that ``black-bounce'' models have recently become quite popular, though more typically for bounces back into our own universe, see for instance references~\cite{Barcelo:2014, Barcelo:2014b, Barcelo:2015, Barcelo:2016, Garay:2017, Rovelli:2014, Haggard:2015, Christodoulou:2016, DeLorenzo:2015, Malafarina:2017, Olmedo:2017, Barrau:2018, Malafarina:2018}. Not all of these bounce models are entirely equivalent, either to each other or to the bounce scenarios of this current article.

In this paper we will investigate whether the above mentioned physical scenarios can actually be described by metric (\ref{E:dmetric}) and analyze interesting physical characteristics of this geometry. 
The paper is structured as follows: In Section~\ref{S:basics} we set up the generic geometric basics for our models; 
then in Section~\ref{S:einstein} we discuss the Einstein tensor and related energy conditions. 
In Section~\ref{S:models} we develop some specific physical models (with either ingoing or outgoing null flux), and exhibit some relevant Carter--Penrose diagrams. 
We discuss the overall framework and draw conclusions in Section~\ref{S:discussion}. Specific and exhaustive technical computations of curvature tensors and curvature invariants are relegated to the appendix.

\section{Geometric basics}\label{S:basics}
In the present work we consider a coordinate patch in which the metric takes the form
\begin{equation}\label{E:VBBmetric}
ds^{2}=-\left(1-\frac{2m(w)}{\sqrt{r^{2}+a^{2}}}\right)dw^{2}-(\pm 2 \,dw \,dr)
+\left(r^{2}+a^{2}\right)\left(d\theta^{2}+\sin^{2}\theta \;d\phi^{2}\right),
\end{equation}
where the coordinates have natural domains:
\begin{equation}
w\in(-\infty,+\infty);\qquad
r\in(-\infty,+\infty);\qquad 
\theta\in [0,\pi];\qquad
\phi\in(-\pi,\pi].
\end{equation}
Here the coordinate $w=\{u,v\}$ denotes what for the region $r>\sqrt{\max\{0,4m^2-a^2\}}$ is \emph{manifestly} an $\{outgoing, ingoing\}$ null time coordinate, thus corresponding respectively to $\{retarded$, $advanced\}$ time, and $\pm\to+$ for $u$, while $\pm\to-$ for $v$.
In the region $r\leq\sqrt{\max\{0,4m^2-a^2\}}$ we continue to use the same nomenclature for the  $w=\{u,v\}$ coordinates.

Note that the same sort of technical issue regarding the precise designation of  $\{outgoing$, $ingoing\}$ coordinates, and $\{retarded,advanced\}$ time,   arise whenever one has multiple domains of outer communication. So even for the maximally extended Schwarzschild spacetime, or the maximally extended Reissner--Nordstr\"om spacetime, one has to define $\{outgoing,ingoing\}$  with respect to a specified asymptotic region --- a specified domain of outer communication.  This technical issue then also afflicts both Morris--Thorne traversable wormholes and the ``black bounces'' of the present article, but does not really require any new physics.

The radial null curves are found by setting 
\begin{equation}
0=ds^2 = dw \left[ \left(1-\frac{2m(w)}{\sqrt{r^{2}+a^{2}}}\right)dw \pm 2\,dr\right], 
\end{equation}
corresponding to
\begin{equation}
dw=0 \qquad \hbox{and} \qquad  dr = \mp {1\over 2} \left(1-\frac{2m(w)}{\sqrt{r^{2}+a^{2}}}\right)dw,
 \end{equation}
 and the associated radial null vectors are proportional to 
\begin{equation}
k^a = (0,\pm 1,0,0) \qquad \hbox{and} \qquad k^a = \left( 1, \mp{1\over2}\left(1-\frac{2m(w)}{\sqrt{r^{2}+a^{2}}}\right); 0,0\right),
\end{equation}
respectively. 

\noindent
That is, for $outgoing$ null coordinates ($retarded$ time) the two radial null vectors are
\begin{equation}
k^a = (0,1,0,0) \qquad \hbox{and} \qquad k^a = \left( 1, -{1\over2}\left(1-\frac{2m(w)}{\sqrt{r^{2}+a^{2}}}\right); 0,0\right).
\end{equation}
Note that in these coordinates the $r$ components of the null vectors $k^a$, that is the $k^r$, are of opposite sign $(+,-)$ in the ``normal region'' $r^2 > 4 m^2-a^2$, but they have the same sign $(+,+)$ between any horizons that may be present $r^2< 4m^2-a^2$. 

In contrast for $ingoing$ null coordinates ($advanced$ time) the two radial null vectors are
\begin{equation}
k^a = (0,-1,0,0) \qquad \hbox{and} \qquad k^a = \left( 1, {1\over2}\left(1-\frac{2m(w)}{\sqrt{r^{2}+a^{2}}}\right); 0,0\right).
\end{equation}
So in these coordinates the $r$ components of the null vectors $k^a$, that is the $k^r$, are of opposite sign $(-,+)$ in the ``normal region'' $r^2 > 4 m^2-a^2$, but they have the same sign $(-,-)$ between any horizons that may be present $r^2< 4m^2-a^2$.

\clearpage
As for the static case analyzed in reference~\cite{black-bounce}, we can define a (radial) ``coordinate speed of light'':
\begin{equation}
c_{radial} = {{\rm d}r\over{\rm d}w} = \mp{1\over2}\left(1-\frac{2m(w)}{\sqrt{r^{2}+a^{2}}}\right).
\end{equation}
If  $2m(w)>a$, this radial  ``coordinate speed of light'' vanishes at
\begin{equation}
r_{AH}(w)=\pm\sqrt{(2m(w))^2-a^2},
\end{equation} so we have a dynamical apparent horizon.

In contrast, for tangential null curves (that is, $dr=0$) we can without any loss of generality set $\phi=0$ and concentrate on
\begin{equation}
0=ds^2 = - \left(1-\frac{2m(w)}{\sqrt{r^{2}+a^{2}}}\right)dw^2 +  (r^2+a^2) d\theta^2, 
\end{equation}
for which the associated tangential null vectors, (defined only for $\sqrt{r^2+a^2}\geq 2 m(w)$), are proportional to 
\begin{equation}
 k^a = \left( \sqrt{r^2+a^2},0; \sqrt{1-\frac{2m(w)}{\sqrt{r^{2}+a^{2}}}} ,0\right).
\end{equation}
We can if desired define a (tangential) ``coordinate speed of light'',
\begin{equation}
c_{tangential} = \sqrt{r^2+ a^2}\;  {{\rm d}\theta\over{\rm d}w} =  \sqrt{r^2+ a^2}\;  {k^\theta\over k^w} = \sqrt{1-\frac{2m(w)}{\sqrt{r^{2}+a^{2}}}},
\end{equation}
but this quantity is not particularly useful for characterizing the presence of horizons. (In fact $d\theta/d w \to 0$ as one approaches the apparent horizon from large $|r|$, and is undefined for small $|r|$.)

The existence of a future/past event horizon depends on the presence or absence of an apparent horizon in the limit $t\rightarrow\pm\infty$, that is, event horizon existence depends on whether the limit $2m(\pm\infty)/a$ exceeds, equals, or is less than unity.
We already know from the static case~\cite{black-bounce}, that there is a throat/bounce hypersurface at $r=0$. At this hypersurface the induced 3-metric is
\begin{equation}\label{E:VBBthroat}
ds|_\Sigma^{2}=-\left(1-\frac{2m(w)}{a}\right)dw^{2}
+a^{2}\left(d\theta^{2}+\sin^{2}\theta \;d\phi^{2}\right).
\end{equation}

Geometrically, this induced 3-geometry is always a cylinder, though potentially of variable signature.
Specifically this $r=0$ hypersurface is timelike if $2m(w)/a<1$, null (lightlike) if $2m(w)/a=1$, and spacelike if $2m(w)/a>1$. These correspond to a traversable wormhole throat, a one-way null throat, or a ``black-bounce'' respectively, where now (as opposed to the static discussion of reference~\cite{black-bounce}) the nature of the throat can change in a $w$-dependent manner.
Because of this feature, the relevant Carter--Penrose diagrams will thus depend on the entire history of the ratio $2m(w)/a$ over the entire domain $w\in(-\infty,+\infty)$. Since the Carter--Penrose diagrams are constructed to exhibit intrinsically global causal structure, to determine them one needs global information regarding $2m(w)/a$.

\section{Einstein tensor and energy conditions}\label{S:einstein}

In Eddington--Finkelstein coordinates, as long as $a\neq 0$, both the metric $g_{ab}$ and the inverse metric $g^{ab}$ have finite components for all values of $r$. 
Moreover, as was shown in detail for the static case~\cite{black-bounce}, and as we shall analyze for the dynamical case in the appendix below, all the curvature tensors (Riemann, Weyl, Ricci, Einstein) have finite components for all values of $r$.
Consequently, even for a time-dependent $m(w)$ one still has a regular spacetime geometry --- there are no curvature singularities.

We discuss here in some detail the results for the Einstein tensor, since it is strongly related with the stress-energy tensor in GR. 
The Einstein tensor has non-zero components:
\begin{eqnarray}
G_{ww} &=& \mp{2r\;\dot m(w)\over (r^2+a^2)^{3/2}} -
{a^2\left\{1-{2m(w)\over\sqrt{r^2+a^2}}\right\} \left\{ 1-{4m(w)\over\sqrt{r^2+a^2}}\right\} \over(r^2+a^2)^{2} };\\
G_{wr} &=&
\mp a^2 {\sqrt{r^2+a^2}-4m(w)\over(r^2+a^2)^{5/2}};\\
G_{rr} &=& {-2a^2 \over(r^2+a^2)^2};\\
G_{\theta\theta} &=& 
+{a^2(\sqrt{r^2+a^2}-m(w))\over (r^2+a^2)^{3/2}}
= {G_{\phi\phi}\over \sin^2\theta}.
\end{eqnarray}
with $\dot m(w)=dm/dw$. Note that the derivative term $\dot m(w)$  only shows up linearly, and only in a very restricted way. 
In fact we can write
\begin{equation}\label{Gs}
G_{ab} = G_{ab}^{nonderivative} \mp{2r\;\dot m(w)\over (r^2+a^2)^{3/2}} \; (dw)_a (dw)_b.
\end{equation}
where we remind the reader that the upper $-$ sign corresponds to the outgoing coordinate $u$ and the lower $+$ sign
to the ingoing coordinate $v$.
It is interesting to underline that the derivative term is precisely the only term present in the pure Vaidya case where $a=0$. Note that $G_{ab}\propto T_{ab}$. So, it is like we were considering a flux equivalent to that of the Vaidya geometry on top of the (now dynamical) fluid that generates the static spacetime. 
It is in this sense that we will discuss the existence of a null flux proportional to $\dot m(w)$ in the dynamical region of the geometry in Section \ref{S:models}.

Now, let us consider the nature of the matter content generating these geometries. 
We already know that the material supporting the static geometry, with $m(w)=m$, violates the Null Energy Condition (NEC)~\cite{black-bounce}. 
This condition is a necessary requirement for forcing all timelike observers to see non-negative energy densities.
As the NEC is used in the singularity theorems to assure convergence of geodesics in GR, one should already expect to have some violations in wormholes, where the throat has to flare out, or in black bounces, which avoid the formation of singularities~\cite{twisted,Kar:2004, Molina-Paris:1998, Visser:cosmo1999, Barcelo:2000b, Visser:1999-super, Visser:1998-super, Abreu:2008, Abreu:2010, LNP, Martin-Moruno:2013a, Martin-Moruno:2013b,Martin-Moruno:2015, Visser:1994, Visser:1996a, Visser:1996b, Visser:1997-ec}.

In the dynamical case, some results of the static geometry will be recovered, but there will also be some crucial differences. For the specific radial null vector \,$k^a = (0,1,0,0)$\, we have
\begin{equation}\label{NEC1}
T_{ab} k^a k^b \; \propto \; G_{ab} k^a k^b = G_{rr} = - {2a^2\over(r^2+a^2)^2}. 
\end{equation}
This implies that in GR the stress-energy tensor is always NEC violating.
Although the result above is already enough to conclude the violation of the NEC, let us study other contractions in order to figure out the effect of having a non-constant mass. For the other radial null vector $k^a = \left( 1, \mp {1\over2}\left(1-\frac{2m(w)}{\sqrt{r^{2}+a^{2}}}\right) , 0,0\right)$, where the minus sign corresponds to $u$ and the plus sign to $v$, we have
\begin{equation}
G_{ab} k^a k^b = - {a^2\left(\sqrt{r^2+a^2}-2m(w)\right)^2\over2(r^2+a^2)^3} \mp {2r\dot m(w)\over(r^2+a^2)^{3/2}}. 
\end{equation}
The non-derivative term is always NEC violating. The derivative term $\dot m(w)$ might or might not be NEC violating depending on sign. 
When considering ingoing radiation (described by $v$) the stress-energy tensor that can be constructed considering only the derivative term satisfies the NEC for non-decreasing $m(v)$. For outgoing radiation the situation is the opposite, so the NEC is satisfied by that flux for $\dot m(u)<0$.
Overall NEC violation in this particular direction would depend on relative magnitudes and signs.

In contrast, for the transverse null vector $k^a = \left( \sqrt{r^2+a^2},0, \sqrt{1-\frac{2m(w)}{\sqrt{r^{2}+a^{2}}}} ,0\right)$ we have
\begin{equation}
G_{ab} k^a k^b =   {3m(w)a^2(\sqrt{r^2+a^2}-2m(w))\over(r^2+a^2)^2}\mp {2r\dot m(w)\over\sqrt{r^2+a^2}}.
\end{equation}
The non-derivative term is now NEC satisfying for wormholes and outside the horizon of regular black holes. The derivative term $\dot m(w)$ might or might not be NEC violating depending on sign. Overall NEC violation in this particular direction would depend on relative magnitudes and signs. 
However, we emphasize that to violate the NEC it is sufficient  to have even one direction in which we have non-positive contraction $G_{ab} k^a k^b$. This certainly occurs for the radial direction, see (\ref{NEC1}). 

Summarizing, we can write
\begin{equation}
T_{ab} = T_{ab}^{nonderivative}+ T_{ab}^{derivative},\quad {\rm with}\quad T_{ab}^{derivative}\propto \mp\dot m(w)\; (dw)_a (dw)_b.
\end{equation}
Whereas $T_{ab}^{nonderivative}$ always violates the NEC in the radial direction; the flux described by $T_{ab}^{derivative}$ satisfies the NEC for ingoing radiation with $\dot m(v)\geq0$ and for outgoing radiation with  $\dot m(u)\leq 0$.

\section{Physical models}\label{S:models}
In this section we analyze some particular evolutionary scenarios that can be described by the spacetime metric (\ref{E:dmetric}). In particular, we focus on several situations of direct physical interest first taking ingoing Eddington--Finkelstein coordinates and later outgoing Eddington--Finkelstein coordinates. We classify those scenarios as having ingoing or outgoing radiation, respectively, focusing attention on the $T_{ab}^{derivative}$ part of the stress-energy tensor, which is not present in the static case.

Given the fact that different values for our (hypothetical) parameter $a$ correspond to qualitatively different spacetime geometries containing different astrophysical objects, (from traversable wormholes to shining stars to black bounces), a completely general analysis of the global causal structure for this metric is not a viable project.  Even when setting our parameter $a=0$ and recovering the standard Vaidya spacetime a specification must be made as to whether we impose an outgoing/ingoing timelike coordinate $w$, and one must also choose a specific form for the mass function $m(w)$  before any conclusions concerning global causal structure can be made. Accordingly, whilst the metric does not permit a completely general analysis of global causal structure, we may investigate various sub-cases by imposing conditions on the form of our mass function $m(w)$ to correspond to specific physical scenarios of interest, and thereby make appropriate conclusions concerning the corresponding global causal structure.

\subsection{Models with ingoing radiation (accretion) }
Let us now focus on the spacetime metric (\ref{E:dmetric}) with ingoing (advanced) Eddington--Finkelstein coordinates. That is
\begin{equation}\label{E:dmetricv}
ds^{2}=-\left(1-\frac{2m(v)}{\sqrt{r^{2}+a^{2}}}\right)dv^{2}+ 2 \,dv \,dr
+\left(r^{2}+a^{2}\right)\left(d\theta^{2}+\sin^{2}\theta \;d\phi^{2}\right).
\end{equation}
As is well known, in the standard Vaidya situation~\cite{Vaidya:1951a,Vaidya:1951b,Vaidya:1999a,Vaidya:1999b,Vaidya:1970, Wang:1998, Parikh:1998}, (that is for $a=0$), this metric describes an ingoing null flux with $T_{vv}\propto 2\dot m(v)/r^2$. So, the black hole mass increases as a result of an ingoing flux with positive energy. 
When $a\neq0$, the geometry is generated by a non-vanishing stress-energy tensor even in the static case, $m(v)=m$. But, 
as we have discussed in the previous section, when one allows $m(v)$ to be a dynamical quantity, then an extra null flux term will appear in that tensor. That is
\begin{equation}
T_{ab}^{derivative}\propto \dot m(v)\; (dv)_a (dv)_b
\end{equation}
So, the derivative contribution to the null flux is positive for $\dot m(v)>0$ and negative for $\dot m(v)<0$.
In this case we can distinguish three different physically relevant situations. Denoting $m_0$ as the initial mass, two of them are characterized by $\dot m(v)>0$ and the last one by $\dot m(v)<0$. These three scenarios are:

\paragraph{Growing black-bounce ($a<2m_0$).} 
For an outside observer in our universe the initial situation will be similar to that for a black hole with an apparent horizon given by $r_{+0}=\sqrt{(2m_0)^2-a^2}$; however, the interior region will instead describe a bounce into another universe. Now, turn on an additional positive ingoing null flux by considering a non-constant increasing function $m(v)$. With the increase of $m(v)$, the radius of the apparent horizon will also increase, $r_+(v)$, leading to a bigger black object.

A particularly simple example is that of piecewise-linear growth, given by
\begin{equation}
m(v)= \left\{ \begin{array}{ll}
             m_0>a/2, &\qquad   v\leq 0; \\
             m_0+\alpha v, &\qquad 0 < v < v_f; \\
             m_f=m_0+\alpha\, v_f, &\qquad v\geq v_f;
             \end{array}
   \right.
\end{equation}
with $\alpha>0$. 
As an astrophysical object this scenario models a nonsingular black hole (with a black bounce at its core) which is accreting ordinary matter over time.

\begin{figure}[!htb]
\begin{center}
\includegraphics[scale=0.90]{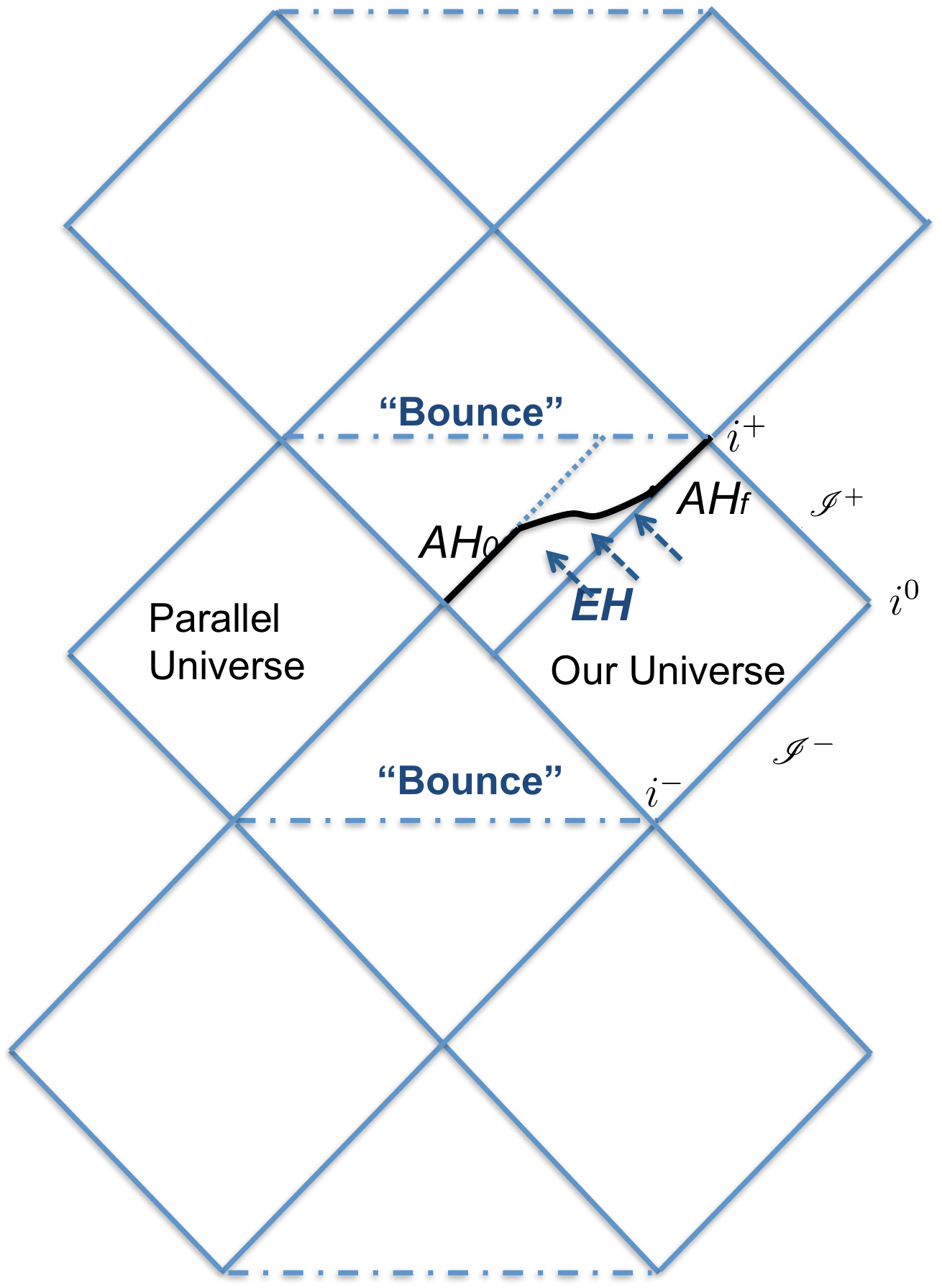}
\end{center}
\caption{Carter--Penrose diagram for a growing black-bounce. There is positive radiation being accreted by the black-bounce for $0 < v < v_f$ (shown by arrows in the diagram). The apparent horizon evolves from $AH_0$ to $AH_f$. Note that before the influx of this radiation the diagram is symmetric; 
however, during accretion of the fluid by the black-bounce the diagram is asymmetric, and after the subsequent post-accretion bounce the diagram is again symmetric but shifted to the right.}
\label{F:1}
\end{figure}

\begin{figure}[!htb]
\begin{center}
\includegraphics[scale=0.90]{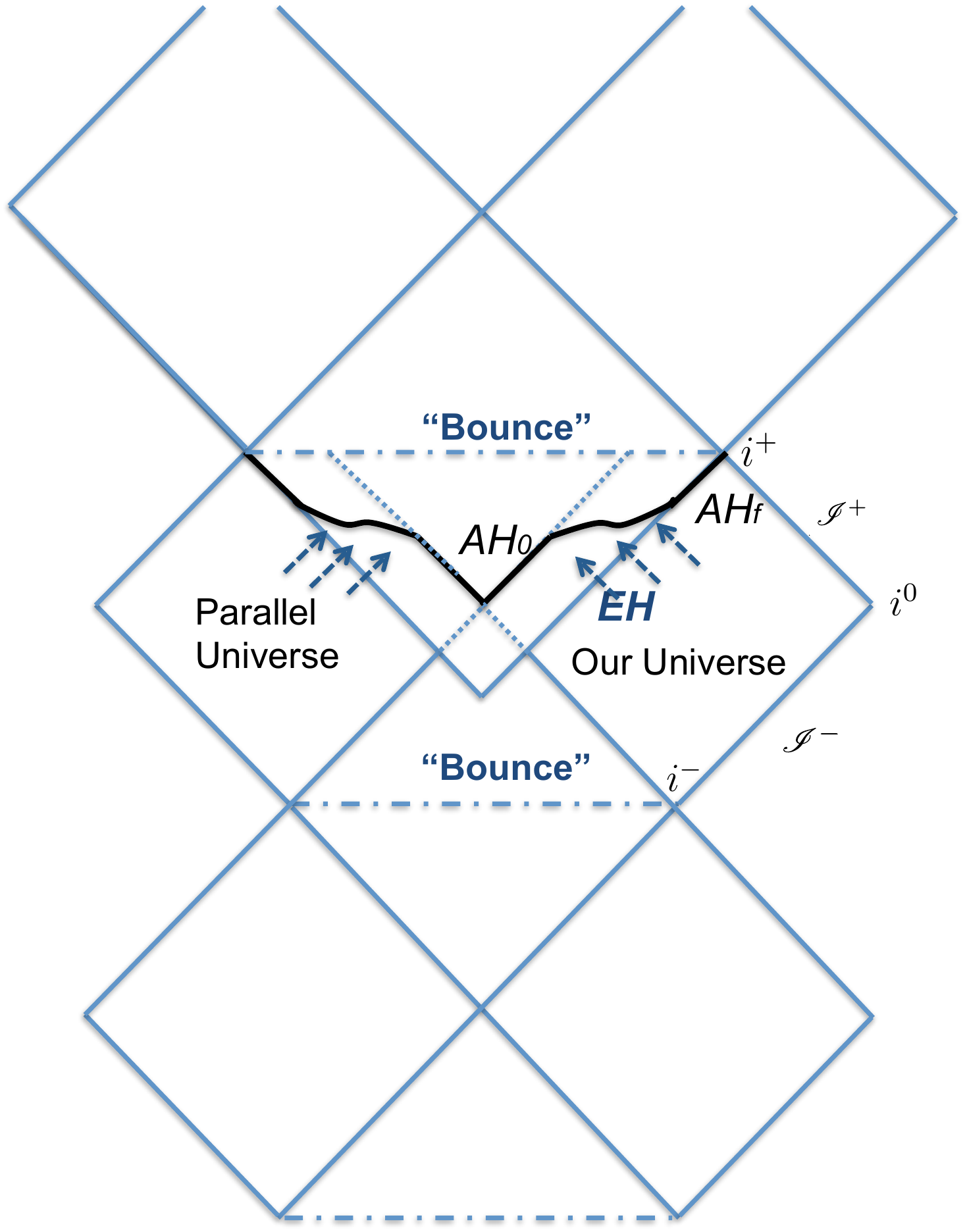}
\end{center}
\caption{Carter--Penrose diagram for a growing black-bounce. We now restore the symmetric character of the diagram by assuming that, for some reason, there is also positive radiation being accreted by the black-bounce of the parallel universe for $0 < v < v_f$.}
\label{F:2}
\end{figure}

In this case, there is an apparent horizon at 
\begin{equation}
r_{+}(v)=\sqrt{(2m(v))^2-a^2}
\end{equation}
and an event horizon, which partially overlaps with the final apparent horizon, located at 
\begin{equation}
r_{+f}=\sqrt{4m_f^2-a^2} = \sqrt{(2m_0+2\alpha\, v_f)^2-a^2}.
\end{equation}
The Carter--Penrose diagram for this scenario can be seen in Figure \ref{F:1}, whereas in Figure \ref{F:2} we show the resulting spacetime if one considers that a similar flux is turned on in the parallel universe.

Note that the choice of a piecewise-linear growth is just for simplicity of exposition. The only real features of $m(v)$ that we are using in constructing the Carter--Penrose diagram are the assumed existence of the limits 
\begin{equation}
m(v\to+\infty) > m( v\to -\infty) > a/2, \qquad \hbox{and the condition} \qquad \dot m(v) \geq 0.
\end{equation}

\clearpage

\begin{figure}[!htb]
\begin{center}
\includegraphics[scale=0.9]{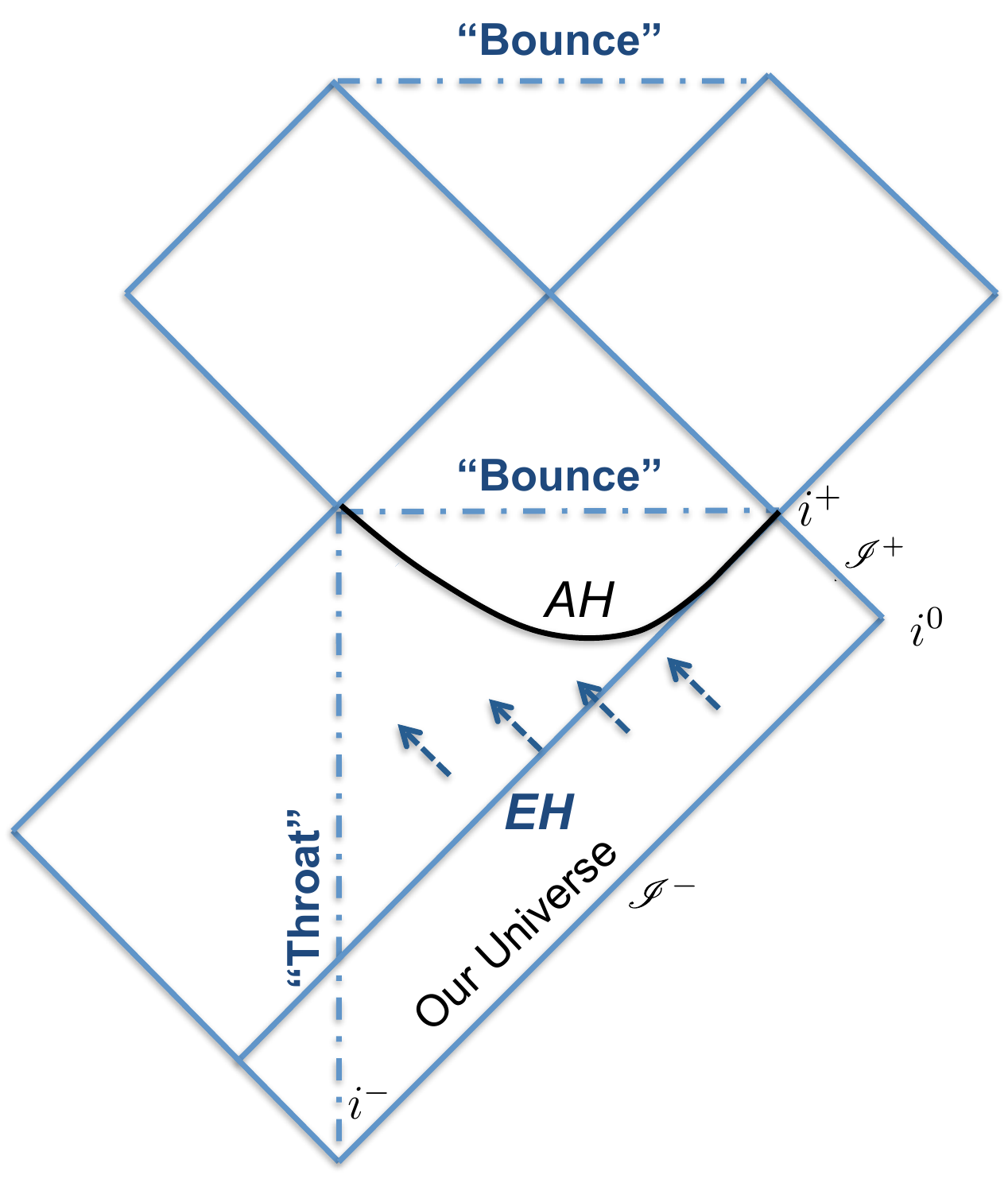}
\end{center}
\caption{Carter--Penrose diagram for a wormhole to black-bounce transition. There is an incoming flux of positive radiation into the wormhole (depicted by arrows) that causes its transmutation into a black-bounce. That is, the timelike wormhole throat hypersurface becomes a spacelike black-bounce hypersurface, passing through being null at the point from which the apparent horizon emerges. Since there is a final apparent horizon, our universe would have an event horizon 
which cannot end at the throat (which is not a boundary of the spacetime) and, therefore, continues through the other universe.}
\label{F:3}
\end{figure}

\paragraph{Wormhole to black-bounce transition ($a>2m_0$).} 
In this case, the initial scenario will be that of a traversable Morris--Thorne wormhole (which could even have $m_0=0$). 
Now, we again turn on an additional ingoing flux with positive energy, by taking a non-constant increasing function $m(v)$. At first, this will have no effect in the causal properties of the geometry. But, if the increasing function $m(v)$ crosses the critical value $a/2$, then we will momentarily have a one-way wormhole, and then a regular black hole will form.  So sufficiently large ingoing positive null flux will lead to the transition from a wormhole to a regular black hole. As in the previous case,  for simplicity of exposition we could consider the piecewise-linear growth function
\begin{equation}
m(v)= \left\{ \begin{array}{ll}
             m_0<a/2, &\qquad   v\leq 0; \\
             m_0+\alpha v, &\qquad 0 < v < v_f; \\
             m_f=m_0+\alpha\, v_f>a/2, &\qquad v\geq v_f.
             \end{array}
   \right.
\end{equation}
The Carter--Penrose diagram of this scenario can be seen in Figure \ref{F:3}. This situation can be interpreted as the accretion of energy satisfying the NEC onto a wormhole. When the mass of the hole reach the value $2m(v)=a$, its causal character changes from timelike to spacelike, momentarily passing through null. At that point, an apparent horizon forms to hide the spacelike bounce. The event horizon of our space, which partially overlaps with the final apparent horizon, is placed at
\begin{equation}
r_{+f}=\sqrt{4m_f^2-a^2} =\sqrt{(2m_0+2\alpha\, v_f)^2-a^2} .
\end{equation} 
Note that the choice of a piecewise-linear growth is just for simplicity of exposition. The only real features of $m(v)$ that we are using in constructing the Carter--Penrose diagram are the assumed existence of the limits 
\begin{equation}
m(v\to+\infty) <a/2, \qquad m( v\to -\infty) > a/2, \quad \hbox{and the condition} \quad \dot m(v) \geq 0.
\end{equation}

\clearpage

\begin{figure}[!htb]
\begin{center}
\includegraphics[scale=0.90]{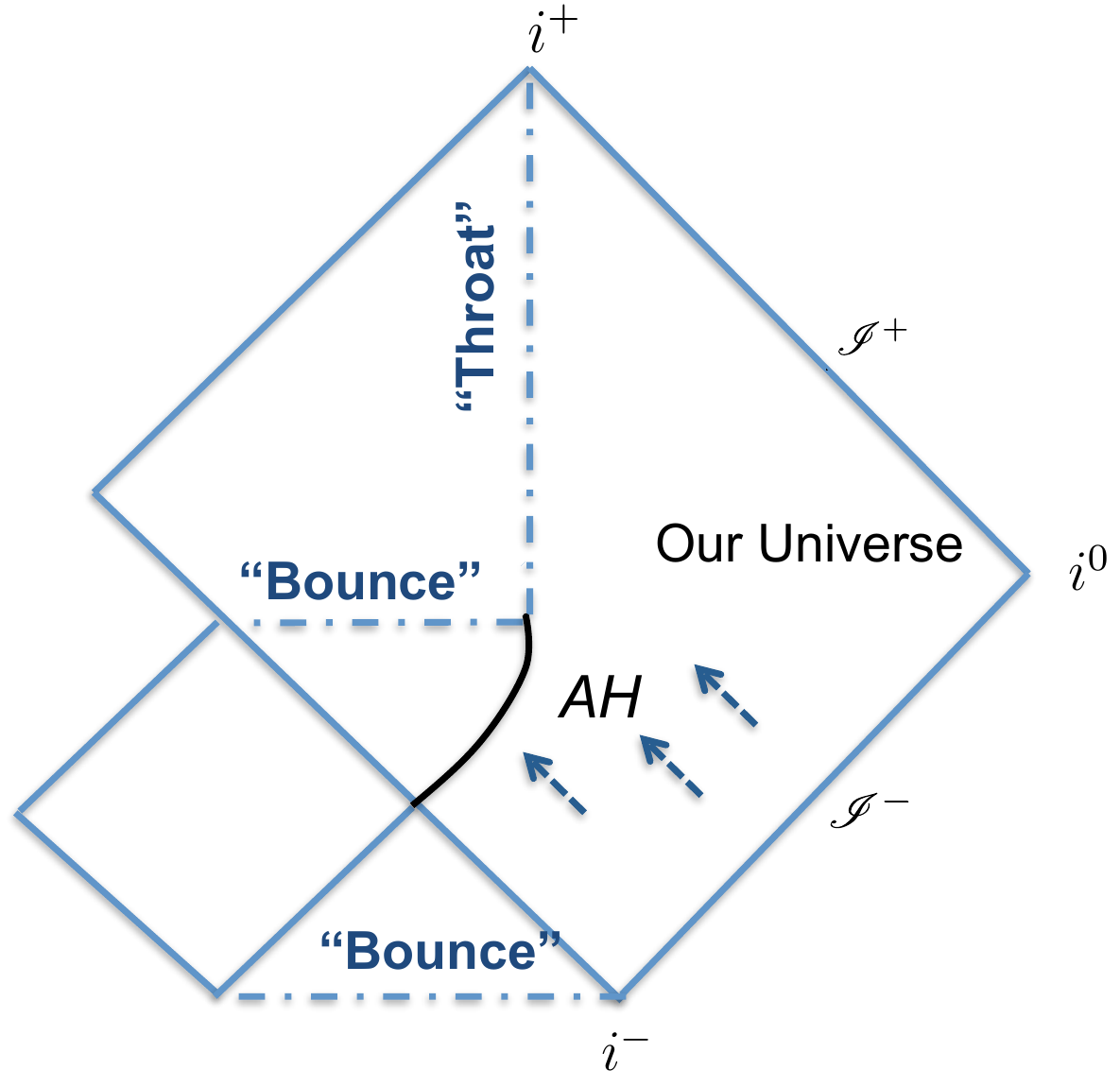}
\qquad
\end{center}
\caption{Carter--Penrose diagram for a black-bounce to wormhole transition due to the accretion of phantom energy. The arrows indicate the region where the phantom fluid is being accreted. There is a black-bounce in our universe, characterized by an apparent horizon, that converts into a wormhole. Therefore, there is no event horizon in our universe.}
\label{F:4}
\end{figure}

\paragraph{Phantom energy accretion onto a black-bounce.}
We could also consider the case in which the additional ingoing flux that we turn on when allowing $m(v)$ to vary is characterized by a negative energy density. This type of exotic fluid is called phantom energy in a cosmological setting. The accretion of phantom energy into black holes has been studied in the test-fluid regime~\cite{Babichev:2004yx,Babichev:2004qp,MartinMoruno:2006mi,GonzalezDiaz:2007gt,Martin-Moruno:2007,Madrid:2010}, predicting a decrease of the black hole mass. With the present formalism we could take into account the back-reaction of this process, by using the advanced metric (\ref{E:dmetricv}), but considering $\dot m(v)<0$. 
However, an important difference with that picture is that our static geometry is a non-vacuum solution of the Einstein equations. We consider again for simplicity a finite region of piecewise-linear evolution, that is now
\begin{equation}
m(v)= \left\{ \begin{array}{ll}
             m_0>a/2, &\qquad   v\leq 0; \\
             m_0-\alpha v, &\qquad 0 < v < v_f; \\
             m_f=m_0-\alpha\, v_f<a/2, &\qquad v\geq v_f;
             \end{array}
   \right.
\end{equation}
The apparent horizon of the regular black hole decreases due to the accretion of phantom energy. At $2m(v)=a$, this horizon disappears and the bounce surface is null, becoming then timelike. So, an ideal observer in this universe will see a black hole that is converted into a wormhole.
The Carter--Penrose diagram of this scenario can be seen in Figure \ref{F:4}.

Note that the choice of a piecewise-linear mass decrease is just for simplicity of exposition. The only real features of $m(v)$ that we are using in constructing the Carter--Penrose diagram are the assumed existence of the limits 
\begin{equation}
m(v\to+\infty) >a/2, \qquad m( v\to -\infty) < a/2, \quad \hbox{and the condition} \quad \dot m(v) \leq 0.
\end{equation}

\clearpage
\subsection{Model with outgoing radiation (evaporation)}
It is also interesting to consider the spacetime metric (\ref{E:dmetric}) with outgoing (retardad) Eddington--Finkelstein coordinates. That is
\begin{equation}\label{E:dmetricu}
ds^{2}=-\left(1-\frac{2m(u)}{\sqrt{r^{2}+a^{2}}}\right)du^{2}- 2 \,du \,dr
+\left(r^{2}+a^{2}\right)\left(d\theta^{2}+\sin^{2}\theta \;d\phi^{2}\right).
\end{equation}
For $a=0$, this is the standard retarded Vaidya metric that describes an outgoing null flux with $T_{uu}\propto -2\dot m(u)/r^2$. 
This scenario can be used to describe classically the back reaction of the semi-classical Hawking radiation by a black hole, in which case there is a positive outgoing flux of radiation that corresponds to a decrease of the black hole mass.
For our case, $a\neq0$ 
and we have a non-vacuum solution even for $m(u)=m$. So, when $m(u)$ varies, an extra null flux term will appear in that tensor (see  section \ref{S:einstein}), with
\begin{equation}
T_{ab}^{derivative}\propto -\dot m(u)\; (du)_a (du)_b .
\end{equation}
Therefore, we have a positive outgoing flux for $\dot m(u)<0$.

\begin{figure}[!htb]
\begin{center}
\includegraphics[scale=0.90]{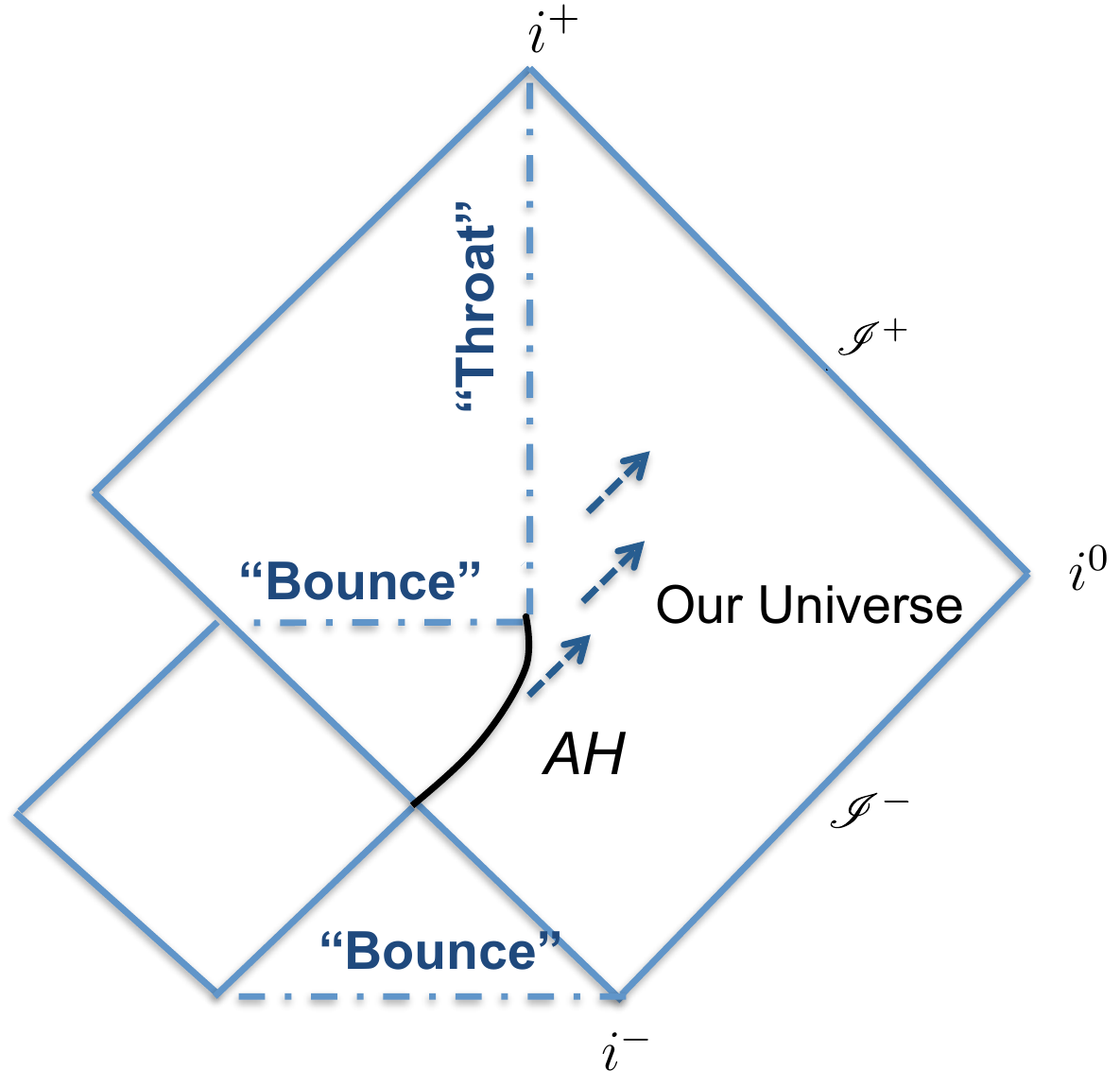}
\qquad
\end{center}
\caption{Carter--Penrose diagram for a black-bounce to wormhole transition due to the emission of positive energy. This diagram is very similar to that shown in Figure~\ref{F:4}, however, now there is a (positive) flux being emitted by the black-bounce and wormhole.}
\label{F:5}
\end{figure}

\paragraph{Classical effective description of black hole radiation.} 
Of course, one should first study carefully the semi-classical properties of this solution to interpret the outgoing flux as semi-classical~\cite{sparsity, sparsity2}. However, it is interesting to consider this scenario as we may have a black-bounce to wormhole transition similar to that already considered in the previous subsection. In this case it would be interesting to emphasize that the remnant of the black-bounce would be a wormhole.
The Carter--Penrose diagram of this scenario can be seen in Figure \ref{F:5}.

\clearpage

\section{Discussion}\label{S:discussion}

In this article we have presented several simple and tractable scenarios for the time evolution of the regular ``black-bounce''/traversable wormhole spacetime considered in reference~\cite{black-bounce}. These models provide a good framework for considering ``black-bounce'' $\longleftrightarrow$ traversable wormhole transitions.
However, despite the generality of our simple models, it should be noted that in this framework a black-bounce cannot be formed by gravitational collapse from an ordinary stellar object. This is because in the limit $m\rightarrow0$, we have a traversable wormhole instead of Minkowski spacetime. So, in order to describe the physically relevant situation of stellar collapse one should go beyond our simple treatment above and consider \emph{both} $a(w)$ and $m(w)$ appropriately. Note that computations would then be significantly more complex, and more importantly that there would then be a \emph{qualitative} difference between the cases $a=0$ and $a\neq 0$.  We leave such considerations for future work.

\section*{Acknowledgments}
PMM acknowledges financial support from the project FIS2016-78859-P (AEI/FEDER, UE).
MV acknowledges direct financial support via the Marsden Fund administered by the Royal Society of New Zealand.
AS acknowledges indirect financial support via the Marsden Fund administered by the Royal Society of New Zealand.

\appendix

\section*{Appendix: Curvature tensors and curvature invariants}\label{S:appendix}
\addcontentsline{toc}{section}{Appendix: Curvature tensors and curvature invariants}
\def\thesection{A} 
The key point is that in Eddington--Finkelstein coordinates, as long as $a\neq 0$, both the metric $g_{ab}$ and the inverse metric $g^{ab}$ have finite components for all values of $r$. 
Specifically (taking upper sign for $u$, lower sign for $v$)
\begin{equation}
g_{ab} = \left[ \begin{array}{crcc} -\left(1-{2m(w)\over\sqrt{r^2+a^2}} \right)&  \mp1 & 0 & 0\\
\mp1 &\;0 & 0 & 0\\
0 & 0 & (r^2+a^2) & 0 \\
0 & 0 & 0 & (r^2+a^2)\sin^2\theta\\
  \end{array}\right]
\end{equation}
and 
\begin{equation}
g^{ab} = \left[ \begin{array}{rccc}0&  \mp1 & 0 & 0\\
\mp1 &  \;+\left(1-{2m(w)\over\sqrt{r^2+a^2}} \right) & 0 & 0\\
0 & 0 & {1\over(r^2+a^2)} & 0 \\
0 & 0 & 0 & {1\over(r^2+a^2)\sin^2\theta}\\
  \end{array}\right]
\end{equation}
Similarly we shall soon see that the curvature tensors (Riemann, Weyl, Ricci, Einstein) have finite components for all values of $r$.
Consequently, even for a time-dependent $m(w)$ one still has a regular spacetime geometry --- there are no curvature singularities.

With this in mind, for simplicity we first consider the non-zero components of the Weyl tensor:
\begin{eqnarray}
C_{wrwr} &=& {m(w) ( a^2-2r^2) \over(r^2+a^2)^{5/2}} - {2a^2 \over3(r^2+a^2)^2}\nonumber\\
 &=& \mp{2C_{w\theta r\theta} \over r^2+a^2}   =  \mp{2  C_{w\phi r\phi} \over (r^2+a^2)\sin^2\theta } 
 =  -{C_{\theta\phi \theta\phi}\over (r^2+a^2)^2\sin^2\theta } \; ;\qquad
 \\
 C_{w\theta w\theta} &=&
    -{(2r^2-a^2) m(w)^2\over (r^2+a^2)^2}  + {(6r^2-7a^2)m(w)\over6(r^2+a^2)^{3/2}} + {a^2\over 3(r^2+a^2) } 
    = {C_{w\phi w\phi} \over \sin^2\theta}.
\end{eqnarray}
Note that there are no derivative contributions (no $\dot m(w) = dm(w)/dw$ contributions) to the Weyl tensor, and that the Weyl tensor components are finite at all values of $r$.

For the Riemann tensor the non-zero components are a little more complicated:
\begin{eqnarray}
R_{wrwr} &=&-  {(2r^2-a^2)m(w)\over(r^2+a^2)^{5/2}};\\
R_{w\theta r\theta} &=&\pm {r^2m(w)\over(r^2+a^2)^{3/2}} = {R_{w\phi r\phi}\over \sin^2\theta};\\
R_{r\theta r\theta} &=&- {a^2\over(r^2+a^2)} = {R_{r\phi r\phi}\over \sin^2\theta};\\
 R_{\theta\phi\theta\phi}&=& \left({2r^2 m(w)\over \sqrt{r^2+a^2}} + a^2 \right) \sin^2\theta;\\
 R_{w\theta w\theta} &=&  \mp {r \; \dot m(w)\over \sqrt{r^2+a^2}} + {r^2 m(w)\over (r^2+a^2)^{3/2}} - {2 r^2 m(w)^2 \over (r^2+a^2)^2} 
 = {R_{w\phi w\phi} \over \sin^2\theta}.
\end{eqnarray}
Note that the derivative term $\dot m(w)$  only shows up linearly, and only in a very restricted way. 
The Ricci tensor has non-zero  components:
\begin{eqnarray}
R_{uu} &=& \mp{2r\;\dot m(w)\over (r^2+a^2)^{3/2}} + {a^2m(w)\; \left\{1-{2m(w)\over\sqrt{r^2+a^2}}\right\} \over(r^2+a^2)^{5/2}};\\
R_{ur} &=&\pm{a^2 m(w)\over (r^2+a^2)^{5/2}};\\
R_{rr} &=& {-2a^2 \over(r^2+a^2)^2};\\
R_{\theta\theta} &=& {2 a^2 m(w)\over (r^2+a^2)^{3/2}} = {R_{\phi\phi}\over \sin^2\theta}.
\end{eqnarray}
Note that the derivative term $\dot m(w)$  only shows up linearly, and only in a very restricted way. 
In fact for the outgoing $u$ coordinate we we can write
\begin{equation}
R_{ab} = R_{ab}^{nonderivative} -{2r\;\dot m(u)\over (r^2+a^2)^{3/2}} \; (du)_a (du)_b.
\end{equation}
On the other hand, if we had taken instead the ingoing coordinate $v$, we would have obtain $R_{vr}=-R_{ur}$ and a sign flip in the derivative term of $R_{vv}$ with respect that of $R_{uu}$. That is,
\begin{equation}
R_{ab} = R_{ab}^{nonderivative} +{2r\;\dot m(v)\over (r^2+a^2)^{3/2}} \; (dv)_a (dv)_b.
\end{equation}
The Einstein tensor has been discussed in Section~\ref{S:einstein}, and those formulae will not be repeated here. 

Note that all of these curvature tensor components are finite at all values of $r$. 
From the discussion above, it is already clear that all of the (polynomial) curvature invariants are all finite for all values of $r$.
For instance, the Ricci scalar is:
\begin{equation}
    R = - {2a^2\over(r^2+a^2)^2} \left\{ 1 - {3\,m(w)\over\sqrt{r^2+a^2}}\right\}.
\end{equation}
Note this is independent of the derivative term $\dot m(w)$.  

Furthermore, the Ricci contraction $R_{ab}R^{ab}$ is:
\begin{eqnarray}
R_{ab}R^{ab} &=& \pm{8a^2r \; \dot m(w) \over(r^2+a^2)^{7/2} } 
+ {4a^4\over (r^2+a^2)^4} \left\{1 - {3m(w)\over\sqrt{r^2+a^2}} + {9 m(w)^2\over2(r^2+a^2)}\right\};\\
&=& \pm{8a^2r \; \dot m(w) \over(r^2+a^2)^{7/2} } 
+ {4a^4\over (r^2+a^2)^4} \left\{ \left(1 - {3m(w)\over2\sqrt{r^2+a^2}}\right)^2 + {9 m(w)^2\over4(r^2+a^2)}\right\}.\qquad
\end{eqnarray}
Note that the derivative term $\dot m(w)$  only shows up linearly. 
Note that the non-derivative contribution is a sum of squares and so automatically non-negative. In 3+1 dimensions $G_{ab} G^{ab} = R_{ab} R^{ab}$, so the  $G_{ab} G^{ab}$ contraction provides nothing new.

The Weyl contraction $C_{abcd}C^{abcd}$ is a perfect square
\begin{equation}
    C_{abcd}C^{abcd} = {16a^4\over(r^2+a^2)^4} \left\{1-{3m(w)\over2\sqrt{r^2+a^2}} + {3m(w)r^2\over a^2\sqrt{r^2+a^2}} \right\}^2.
\end{equation}

The Kretschmann scalar is:
\begin{equation}
R_{abcd} \, R^{abcd} = C_{abcd} \, C^{abcd} +2 R_{ab}\, R^{ab} - \frac{1}{3}R^2,
\end{equation}
and so (in view of the above) without further calculation we have
\begin{eqnarray}
R_{abcd}\,R^{abcd} &=& \pm{16a^2r \; \dot m(w) \over(r^2+a^2)^{7/2} }\\
&& +{12a^4\over (r^2+a^2)^4} 
\left\{1+ {8m(w)(r^2-a^2)\over3a^2\sqrt{r^2+a^2}} + {m(w)^2(4r^4-4a^2r^2+3a^4)\over a^4(r^2+a^2)}\right\}.
\nonumber
\end{eqnarray}
Note that the derivative term $\dot m(w)$  only shows up linearly.
All the curvature invariants are well-behaved everywhere throughout the spacetime.



\begin{thebibliography}{69}  


\bibitem{Bardeen:1968}
J.~M.~Bardeen, ``Non-singular general-relativistic gravitational collapse'', \\
Proceedings of International Conference GR5, 1968, Tbilisi, USSR, p. 174.

\bibitem{Bergmann-Roman}
Thomas A. Roman and Peter G. Bergmann, 
``Stellar collapse without singularities?'',\\
Phys. Rev. D {\bf28} (1983) 1265--1277.\\
doi: \url{https://doi.org/10.1103/PhysRevD.28.1265}

\bibitem{Hayward:2005}
  S.~A.~Hayward,
  ``Formation and evaporation of regular black holes'',\\
  Phys.\ Rev.\ Lett.\  {\bf 96} (2006) 031103\\
  doi: \url{https://doi.org/10.1103/PhysRevLett.96.031103}
  [gr-qc/0506126].
  
\bibitem{Bardeen:2014}
  J.~M.~Bardeen,
  ``Black hole evaporation without an event horizon'',
  arXiv:1406.4098 [gr-qc].
  
  \bibitem{Frolov:2014} 
  V.~P.~Frolov,
  ``Information loss problem and a black hole model with a closed apparent horizon'',
  JHEP {\bf 1405}, 049 (2014)\\
  doi: \url{https://doi.org/10.1007/JHEP05(2014)049}
  [arXiv:1402.5446 [hep-th]].
  

 \bibitem{Frolov:2014b}
  V.~P.~Frolov,
  ``Do Black Holes Exist?'',
  arXiv:1411.6981 [hep-th].
 
 \bibitem{Frolov:2016}
  V.~P.~Frolov,
  ``Notes on nonsingular models of black holes'',\\
  Phys.\ Rev.\ D {\bf 94} (2016) no.10,  104056\\
  doi: \url{https://doi.org/10.1103/PhysRevD.94.104056}
  [arXiv:1609.01758 [gr-qc]].
 
 \bibitem{Frolov:2017}
  V.~P.~Frolov and A.~Zelnikov,
  ``Quantum radiation from an evaporating nonsingular black hole'',\\
  Phys.\ Rev.\ D {\bf 95} (2017) no.12,  124028\\
  \url{https://doi.org/doi:10.1103/PhysRevD.95.124028}
  [arXiv:1704.03043 [hep-th]].

  
  \bibitem{Frolov:2018}
  V.~P.~Frolov,
  ``Remarks on non-singular black holes'',\\
  EPJ Web Conf.\  {\bf 168} (2018) 01001
  doi: \url{https://doi.org/10.1051/epjconf/201816801001}
  [arXiv:1708.04698 [gr-qc]].
  
  \bibitem{Cano:2018}
  P.~A.~Cano, S.~Chimento, T.~Ort\'in and A.~Ruip\'erez,
  ``Regular Stringy Black Holes?'',\\
  arXiv:1806.08377 [hep-th].
  
  \bibitem{Bardeen:2018}
  J.~M.~Bardeen,\\
  ``Models for the nonsingular transition of an evaporating black hole into a white hole'',\\
  arXiv:1811.06683 [gr-qc].
  
  \bibitem{regular}
  R.~Carballo-Rubio, F.~Di Filippo, S.~Liberati, C.~Pacilio and M.~Visser,\\
  ``On the viability of regular black holes'',
  JHEP {\bf 1807} (2018) 023
  doi: \url{https://doi.org/10.1007/JHEP07(2018)023}
  [arXiv:1805.02675 [gr-qc]].
  
 \bibitem{beyond}
  R.~Carballo-Rubio, F.~Di Filippo, S.~Liberati and M.~Visser,\\
  ``Phenomenological aspects of black holes beyond general relativity'',\\
  Physical Review D {\bf98} (2018) 124009.\\
  doi: \url{https://doi.org/10.1103/PhysRevD.98.124009} [arXiv:1809.08238 [gr-qc]].
  
 \bibitem{black-bounce}
 A.~Simpson and M.~Visser,
  ``Black-bounce to traversable wormhole'',\\
   JCAP02(2019)042 
   doi: \url{https://doi.org/10.1088/1475-7516/2019/02/042}
   [arXiv:1812.07114 [gr-qc]].


\bibitem{Morris:1988a}
  M.~S.~Morris and K.~S.~Thorne,
  ``Wormholes in space-time and their use for interstellar travel: A tool for teaching general relativity'',\\
  Am.\ J.\ Phys.\  {\bf 56} (1988) 395.
  doi: \url{https://doi.org/10.1119/1.15620}
  

\bibitem{Morris:1988b}
  M.~S.~Morris, K.~S.~Thorne and U.~Yurtsever,\\
  ``Wormholes, Time Machines, and the Weak Energy Condition'',\\
  Phys.\ Rev.\ Lett.\  {\bf 61} (1988) 1446.\\
  doi: \url{https://doi.org/10.1103/PhysRevLett.61.1446}
  
  \bibitem{Visser:1989a}
  M.~Visser,
  ``Traversable wormholes: Some simple examples'',\\
  Phys.\ Rev.\ D {\bf 39} (1989) 3182
  doi: \url{https://doi.org/10.1103/PhysRevD.39.3182}
  [arXiv:0809.0907 [gr-qc]].
  
\bibitem{Visser:1989b}
  M.~Visser,
  ``Traversable wormholes from surgically modified Schwarzschild space-times'',
  Nucl.\ Phys.\ B {\bf 328} (1989) 203\\
 doi:  \url{https://doi.org/10.1016/0550-3213(89)90100-4}\\{}
  [arXiv:0809.0927 [gr-qc]].
  
 \bibitem{Lorentzian}
  M.~Visser, ``Lorentzian wormholes: From Einstein to Hawking", \\
  (AIP Press, now Springer, New York, 1995).
  
\bibitem{Visser:2003}
  M.~Visser, S.~Kar and N.~Dadhich,\\
  ``Traversable wormholes with arbitrarily small energy condition violations'',\\
  Phys.\ Rev.\ Lett.\  {\bf 90} (2003) 201102\\
  doi: \url{https://doi.org/10.1103/PhysRevLett.90.201102}
  [gr-qc/0301003].
  

\bibitem{Hochberg:1997}
  D.~Hochberg and M.~Visser,\\
  ``Geometric structure of the generic static traversable wormhole throat'',\\
  Phys.\ Rev.\ D {\bf 56} (1997) 4745
  doi: \url{https://doi.org/10.1103/PhysRevD.56.4745}
  [gr-qc/9704082].
  
\bibitem{Poisson:1995}
  E.~Poisson and M.~Visser,
  ``Thin shell wormholes: Linearization stability'',\\
  Phys.\ Rev.\ D {\bf 52} (1995) 7318
  doi: \url{https://doi.org/10.1103/PhysRevD.52.7318}
  [gr-qc/9506083].
  

\bibitem{Barcelo:2000}
  C.~Barcel\'o and M.~Visser,
  ``Scalar fields, energy conditions, and traversable wormholes'',
  Class.\ Quant.\ Grav.\  {\bf 17} (2000) 3843\\
  doi: \url{https://doi.org/10.1088/0264-9381/17/18/318}
  [gr-qc/0003025].
  

\bibitem{Hochberg:1998}
  D.~Hochberg and M.~Visser,
  ``The Null energy condition in dynamic wormholes'',\\
  Phys.\ Rev.\ Lett.\  {\bf 81} (1998) 746\\
  doi: \url{https://doi.org/10.1103/PhysRevLett.81.74}
  [gr-qc/9802048].
  
\bibitem{Cramer:1994}
  J.~G.~Cramer, R.~L.~Forward, M.~S.~Morris, M.~Visser, G.~Benford and G.~A.~Landis,
  ``Natural wormholes as gravitational lenses'',
  Phys.\ Rev.\ D {\bf 51} (1995) 3117\\
  doi: \url{https://doi.org/10.1103/PhysRevD.51.3117}
  [astro-ph/9409051].
  
 
\bibitem{Visser:1997}
  M.~Visser and D.~Hochberg,
  ``Generic wormhole throats'',\\
  Annals Israel Phys.\ Soc.\  {\bf 13} (1997) 249
  [gr-qc/9710001].


 
\bibitem{Barcelo:1999}
  C.~Barcel\'o and M.~Visser,\\
  ``Traversable wormholes from massless conformally coupled scalar fields'',\\
  Phys.\ Lett.\ B {\bf 466} (1999) 127\\
  doi: \url{https://doi.org/10.1016/S0370-2693(99)01117-X}
  [gr-qc/9908029].
  
    
\bibitem{Garcia:2011}
  N.~M.~Garcia, F.~S.~N.~Lobo and M.~Visser,
  ``Generic spherically symmetric dynamic thin-shell traversable wormholes in standard general relativity'',
  Phys.\ Rev.\ D {\bf 86} (2012) 044026
  doi: \url{https://doi.org/10.1103/PhysRevD.86.044026}
  [arXiv:1112.2057 [gr-qc]].
  
\bibitem{Boonserm:2018}
  P.~Boonserm, T.~Ngampitipan, A.~Simpson and M.~Visser,\\
  ``Exponential metric represents a traversable wormhole'',\\
  Phys.\ Rev.\ D {\bf 98} (2018) no.8,  084048\\
  doi: \url{https://doi.org/10.1103/PhysRevD.98.084048}
  [arXiv:1805.03781 [gr-qc]].
  
  
  \bibitem{Lobo:2004}
  F.~S.~N.~Lobo,
  ``Thin shells around traversable wormholes'',
  gr-qc/0401083.
  
  
  \bibitem{Vaidya:1951a}
  P.~Vaidya,
  ``The Gravitational Field of a Radiating Star'',\\
  Proc.\ Natl.\ Inst.\ Sci.\ India A {\bf 33} (1951) 264.
  
  \bibitem{Vaidya:1951b}
  P.~C.~Vaidya,
  ``Nonstatic Solutions of Einstein's Field Equations for Spheres of Fluids Radiating Energy'',
  Phys.\ Rev.\  {\bf 83} (1951) 10.
  doi: \url{https://doi.org/10.1103/PhysRev.83.10}
  
  \bibitem{Vaidya:1999a}
  P.~C.~Vaidya,
  ``The External Field of a Radiating Star in Relativity'',\\
  Gen.\ Rel.\ Grav.\  {\bf 31} (1999) 119.
  doi: \url{https://doi.org/10.1023/A:1018871522880}
  
  \bibitem{Vaidya:1999b}
  P.~C.~Vaidya,
  ``The Gravitational Field of a Radiating Star'',\\
  Gen.\ Rel.\ Grav.\  {\bf 31} (1999) 121.
  doi: \url{https://doi.org/10.1023/A:1018875606950}
  
  \bibitem{Vaidya:1970}
  W.~B.~Bonnor and P.~C.~Vaidya,
  ``Spherically symmetric radiation of charge in Einstein-Maxwell theory'',
  Gen.\ Rel.\ Grav.\  {\bf 1} (1970) 127.\\
  doi: \url{https://doi.org/10.1007/BF00756891}

  \bibitem{Wang:1998}
  A.~Wang and Y.~Wu,
  ``Generalized Vaidya solutions'',\\
  Gen.\ Rel.\ Grav.\  {\bf 31} (1999) 107
  doi: \url{https://doi.org/10.1023/A:1018819521971}
  [gr-qc/9803038].

  \bibitem{Parikh:1998}
  M.~K.~Parikh and F.~Wilczek,
  ``Global structure of evaporating black holes'',\\
  Phys.\ Lett.\ B {\bf 449} (1999) 24\\
  doi: \url{https://doi.org/10.1016/S0370-2693(99)00071-4}
  [gr-qc/9807031].
  
  
\bibitem{Babichev:2004yx}
  E.~Babichev, V.~Dokuchaev and Y.~Eroshenko,
  ``Black hole mass decreasing due to phantom energy accretion'',
  Phys.\ Rev.\ Lett.\  {\bf 93} (2004) 021102\\
  doi: \url{https://doi.org/10.1103/PhysRevLett.93.021102}\\{}
  [gr-qc/0402089].
  
\bibitem{Babichev:2004qp}
  E.~Babichev, V.~Dokuchaev and Y.~Eroshenko,
  ``Dark energy cosmology with generalized linear equation of state'',
  Class.\ Quant.\ Grav.\  {\bf 22} (2005) 143\\
  doi: \url{https://doi.org/10.1088/0264-9381/22/1/010}
  [astro-ph/0407190].
  
\bibitem{MartinMoruno:2006mi}
  P.~Mart\'{\i}n-Moruno, J.~A.~Jim\'enez Madrid and P.~F.~Gonz\'alez-D\'{\i}az,
  ``Will black holes eventually engulf the universe?'',
  Phys.\ Lett.\ B {\bf 640} (2006) 117\\
  doi: \url{https://doi.org/10.1016/j.physletb.2006.07.067}
  [astro-ph/0603761].
  
\bibitem{GonzalezDiaz:2007gt}
  P.~F.~Gonz\'alez-D\'{\i}az and P.~Mart\'{\i}n-Moruno,
  ``Wormholes in the accelerating universe'', \\
  Proceedings of the MG11 Meeting on General Relativity, 2190-2192 (World Scientific, 2008)
  doi: \url{https://doi.org/10.1142/9789812834300_0358}
  arXiv:0704.1731 [astro-ph].
  
  \bibitem{Martin-Moruno:2007}
  P.~Mart\'{\i}n-Moruno,\\
  ``On the formalism of dark energy accretion onto black- and worm-holes'',\\
  Phys.\ Lett.\ B {\bf 659} (2008) 40
  doi: \url{https://doi.org/10.1016/j.physletb.2007.10.083}\\{}
  [arXiv:0709.4410 [astro-ph]].
  
  \bibitem{Madrid:2010}
  J.~A.~Jim\'enez Madrid and P.~Mart\'{\i}n-Moruno,\\
  ``On accretion of dark energy onto black- and worm-holes'',\\
  in {\it Dark energy: theories, developments, and implications}, 215-239\\
   (Nova Publishers, 2010).
  arXiv:1004.1428 [astro-ph.CO].
  
  
  \bibitem{Lobo:2014}
  F.~S.~N.~Lobo, J.~Martinez-Asencio, G.~J.~Olmo and D.~Rubiera-Garcia,\\
  ``Dynamical generation of wormholes with charged fluids in quadratic Palatini gravity,''\\
  Phys.\ Rev.\ D {\bf 90} (2014) no.2,  024033\\
  doi: \url{https://doi.org/10.1103/PhysRevD.90.024033}
  [arXiv:1403.0105 [hep-th]].
  
  \bibitem{Lobo:2013}
  F.~S.~N.~Lobo, J.~Martinez-Asencio, G.~J.~Olmo and D.~Rubiera-Garcia,\\
  ``Planck scale physics and topology change through an exactly solvable model,''\\
  Phys.\ Lett.\ B {\bf 731} (2014) 163\\
  doi: \url{https://doi.org/10.1016/j.physletb.2014.02.038}
  [arXiv:1311.5712 [hep-th]].
  

  
  
  \bibitem{Barcelo:2014}
  C.~Barcel\'o, R.~Carballo-Rubio, L.~J.~Garay and G.~Jannes,\\
  ``The lifetime problem of evaporating black holes: mutiny or resignation'',\\
  Class.\ Quant.\ Grav.\  {\bf 32} (2015) no.3,  035012\\
  doi: \url{https://doi.org/10.1088/0264-9381/32/3/035012}
  [arXiv:1409.1501 [gr-qc]].
  
 \bibitem{Barcelo:2014b}
  C.~Barcel\'o, R.~Carballo-Rubio and L.~J.~Garay,
  ``Mutiny at the white-hole district'',
  Int.\ J.\ Mod.\ Phys.\ D {\bf 23} (2014) no.12,  1442022\\
  doi: \url{https://doi.org/10.1142/S021827181442022X}
  [arXiv:1407.1391 [gr-qc]].
  
  \bibitem{Barcelo:2015}
  C.~Barcel\'o, R.~Carballo-Rubio, L.~J.~Garay and G.~Jannes,
  ``Do transient white holes have a place in Nature?'',
  J.\ Phys.\ Conf.\ Ser.\  {\bf 600} (2015) no.1,  012033.\\
  doi: \url{https://doi.org/10.1088/1742-6596/600/1/012033}
  
  \bibitem{Barcelo:2016}
  C.~Barcel\'o, R.~Carballo-Rubio and L.~J.~Garay,
  ``Exponential fading to white of black holes in quantum gravity'',
  Class.\ Quant.\ Grav.\  {\bf 34} (2017) no.10,  105007\\
  doi: \url{https://doi.org/10.1088/1361-6382/aa6962}
  [arXiv:1607.03480 [gr-qc]].
  
  \bibitem{Garay:2017}
  L.~J.~Garay, C.~Barcel\'o, R.~Carballo-Rubio and G.~Jannes,
  ``Do stars die too long?'',\\
  doi: \url{https://doi.org/10.1142/9789813226609\_0174}
  
  \bibitem{Rovelli:2014}
  C.~Rovelli and F.~Vidotto,
  ``Planck stars'',
  Int.\ J.\ Mod.\ Phys.\ D {\bf 23} (2014) no.12,  1442026\\
  doi: \url{https://doi.org/10.1142/S0218271814420267}
  [arXiv:1401.6562 [gr-qc]].
  
  \bibitem{Haggard:2015}
  H.~M.~Haggard and C.~Rovelli,
  ``Black to white hole tunneling: An exact classical solution'',\\
  Int.\ J.\ Mod.\ Phys.\ A {\bf 30} (2015) no.28n29,  1545015.\\
  doi: \url{https://doi.org/10.1142/S0217751X15450153}
  
  \bibitem{Christodoulou:2016}
  M.~Christodoulou, C.~Rovelli, S.~Speziale and I.~Vilensky,
  ``Planck star tunneling time: An astrophysically relevant observable from background-free quantum gravity'',
  Phys.\ Rev.\ D {\bf 94} (2016) no.8,  084035
  doi: \url{https://doi.org/10.1103/PhysRevD.94.084035}
  [arXiv:1605.05268 [gr-qc]].
  
  \bibitem{DeLorenzo:2015}
  T.~De Lorenzo and A.~Perez,
  ``Improved Black Hole Fireworks: Asymmetric Black-Hole-to-White-Hole Tunneling Scenario'',
  Phys.\ Rev.\ D {\bf 93} (2016) no.12,  124018\\
  doi: \url{https://doi.org/10.1103/PhysRevD.93.124018}
  [arXiv:1512.04566 [gr-qc]].
  
  \bibitem{Malafarina:2017}
  D.~Malafarina,
  ``Classical collapse to black holes and quantum bounces: A review'',\\
  Universe {\bf 3} (2017) no.2,  48
  doi: \url{https://doi.org/10.3390/universe3020048}
  [arXiv:1703.04138 [gr-qc]].

\bibitem{Olmedo:2017}
  J.~Olmedo, S.~Saini and P.~Singh,
  ``From black holes to white holes: a quantum gravitational, symmetric bounce'',
  Class.\ Quant.\ Grav.\  {\bf 34} (2017) no.22,  225011\\
  doi: \url{https://doi.org/10.1088/1361-6382/aa8da8}
  [arXiv:1707.07333 [gr-qc]].
  
  \bibitem{Barrau:2018}
  A.~Barrau, K.~Martineau and F.~Moulin,
  ``A status report on the phenomenology of black holes in loop quantum gravity: Evaporation, tunneling to white holes, dark matter and gravitational waves'',
  Universe {\bf 4} (2018) no.10,  102\\
  doi: \url{https://doi.org/10.3390/universe4100102}
  [arXiv:1808.08857 [gr-qc]].

  \bibitem{Malafarina:2018}
  D.~Malafarina,
  ``Black Hole Bounces on the Road to Quantum Gravity'',\\
  Universe {\bf 4} (2018) no.9,  92.
  doi: \url{https://doi.org/10.3390/universe4090092}
  
   \bibitem{twisted}
  F.~Gray, J.~Santiago, S.~Schuster and M.~Visser,
  ``Twisted black holes are unphysical'',
  Mod.\ Phys.\ Lett.\ A {\bf 32} (2017) no.18,  1771001\\
  doi: \url{https://doi.org/10.1142/S0217732317710018}
  [arXiv:1610.06135 [gr-qc]].


\bibitem{Kar:2004}
  S.~Kar, N.~Dadhich and M.~Visser,
  ``Quantifying energy condition violations in traversable wormholes'',
  Pramana {\bf 63} (2004) 859\\
  doi: \url{https://doi.org/10.1007/BF02705207}
  [gr-qc/0405103].


\bibitem{Molina-Paris:1998}
  C.~Molina-Par\'is and M.~Visser,
  ``Minimal conditions for the creation of a Friedman-Robertson-Walker universe from a bounce'',
  Phys.\ Lett.\ B {\bf 455} (1999) 90\\
  doi: \url{https://doi.org/10.1016/S0370-2693(99)00469-4}
  [gr-qc/9810023].
  
  \bibitem{Visser:cosmo1999}
  M.~Visser and C.~Barcel\'o,
  ``Energy conditions and their cosmological implications'',\\
  doi: \url{https://doi.org/10.1142/9789812792129\_0014}
  gr-qc/0001099.
  
  \bibitem{Barcelo:2000b}
  C.~Barcel\'o and M.~Visser,
  ``Brane surgery: Energy conditions, traversable wormholes, and voids'',\\
  Nucl.\ Phys.\ B {\bf 584} (2000) 415\\
  doi: \url{https://doi.org/10.1016/S0550-3213(00)00379-5}
  [hep-th/0004022].
  
  \bibitem{Visser:1999-super}
  M.~Visser, B.~Bassett and S.~Liberati,
  ``Perturbative superluminal censorship and the null energy condition'',
  AIP Conf.\ Proc.\  {\bf 493} (1999) no.1,  301\\
  doi: \url{https://doi.org/10.1063/1.1301601}
  [gr-qc/9908023].

  
\bibitem{Visser:1998-super}
  M.~Visser, B.~Bassett and S.~Liberati,
  ``Superluminal censorship'',
  Nucl.\ Phys.\ Proc.\ Suppl.\  {\bf 88} (2000) 267
  doi: \url{https://doi.org/10.1016/S0920-5632(00)00782-9}
  [gr-qc/9810026].
    

\bibitem{Abreu:2008}
  G.~Abreu and M.~Visser,
  ``Quantum Interest in (3+1) dimensional Minkowski space'',\\
  Phys.\ Rev.\ D {\bf 79} (2009) 065004
  doi: \url{https://doi.org/10.1103/PhysRevD.79.065004}
  [arXiv:0808.1931 [gr-qc]].
  



\bibitem{Abreu:2010}
  G.~Abreu and M.~Visser,
  ``The Quantum interest conjecture in (3+1)-dimensional Minkowski space'',
  doi: \url{https://doi.org/10.1142/9789814374552\_0481}
  arXiv:1001.1180 [gr-qc].

\bibitem{LNP}
  P.~Mart\'in--Moruno and M.~Visser,
  ``Classical and semi-classical energy conditions'',\\
  Fundam.\ Theor.\ Phys.\  {\bf 189} (2017) 193 (Lecture Notes in Physics)\\
  doi: \url{https://doi.org/10.1007/978-3-319-55182-1\_9}
  [arXiv:1702.05915 [gr-qc]].
  
 
  

\bibitem{Martin-Moruno:2013a} 
  P.~Mart\'in--Moruno and M.~Visser,
  ``Classical and quantum flux energy conditions for quantum vacuum states'',
  Phys.\ Rev.\ D {\bf 88}, no. 6, 061701 (2013)\\
  doi: \url{https://doi.org/10.1103/PhysRevD.88.061701}\\{}
  [arXiv:1305.1993 [gr-qc]].

\bibitem{Martin-Moruno:2013b}
  P.~Mart\'in--Moruno and M.~Visser,
  ``Semiclassical energy conditions for quantum vacuum states'',
  JHEP {\bf 1309} (2013) 050\\
  doi: \url{https://doi.org/10.1007/JHEP09(2013)050}
  [arXiv:1306.2076 [gr-qc]].
  
  \bibitem{Martin-Moruno:2015}
  P.~Mart\'in--Moruno and M.~Visser,
  ``Semi-classical and nonlinear energy conditions'', \\
  Proceedings of The Fourteenth Marcel Grossmann Meeting, 1442-1447. World Scientific 2017.
  doi: \url{https://doi.org/10.1142/9789813226609\_0126}
  [arXiv:1510.00158 [gr-qc]].

  
  \bibitem{Visser:1994}
  M.~Visser,
  ``Scale anomalies imply violation of the averaged null energy condition'',\\
  Phys.\ Lett.\ B {\bf 349} (1995) 443\\
  doi: \url{https://doi.org/10.1016/0370-2693(95)00303-3}
  [gr-qc/9409043].
 \bibitem{Visser:1996a}
  M.~Visser,
 \leftline{ ``Gravitational vacuum polarization. 1: Energy conditions in the Hartle-Hawking vacuum'',}
  Phys.\ Rev.\ D {\bf 54} (1996) 5103
  doi: \url{https://doi.org/10.1103/PhysRevD.54.5103}
  [gr-qc/9604007].
  
  \bibitem{Visser:1996b}
  M.~Visser,
  ``Gravitational vacuum polarization. 2: Energy conditions in the Boulware vacuum'',
  Phys.\ Rev.\ D {\bf 54} (1996) 5116\\
  doi: \url{https://doi.org/10.1103/PhysRevD.54.5116}
  [gr-qc/9604008].
  
\bibitem{Visser:1997-ec}
  M.~Visser,
  ``Gravitational vacuum polarization. 4: Energy conditions in the Unruh vacuum'',
  Phys.\ Rev.\ D {\bf 56} (1997) 936\\
  doi: \url{https://doi.org/10.1103/PhysRevD.56.936}
  [gr-qc/9703001].

   
 
\bibitem{sparsity}
F.~Gray, S.~Schuster, A.~Van--Brunt and M.~Visser,\\
  ``The Hawking cascade from a black hole is extremely sparse'',\\
  Class.\ Quant.\ Grav.\  {\bf 33} (2016) no.11,  115003\\
  doi: \url{https://doi.org/10.1088/0264-9381/33/11/115003}\\{}
  [arXiv:1506.03975 [gr-qc]].

\bibitem{sparsity2}
M.~Visser, F.~Gray, S.~Schuster and A.~Van--Brunt,\\
  ``Sparsity of the Hawking flux'',\\
  Proceedings of the MG14 Meeting on General Relativity (2017); pp.
  1724-1729\\
  doi: \url{https://doi.org/10.1142/9789813226609\_0175}
  [arXiv:1512.05809 [gr-qc]].

 

  
  
\end{thebibliography}
\end{document}